# Origin of Two Larmor Frequencies in the Coherent Spin Dynamics of Colloidal CdSe Quantum Dots Revealed by Controlled Charging

Rongrong Hu,[†] Dmitri R. Yakovlev,[‡,§,*] Pan Liang,[†] Gang Qiang,[‡] Cong Chen,[†] Tianqing Jia,[†] Zhenrong Sun,[†] Manfred Bayer[‡,§] and Donghai Feng[†,∥,*]

[†] State Key Laboratory of Precision Spectroscopy, East China Normal University, Shanghai 200062, China

[‡] Experimentelle Physik 2, Technische Universität Dortmund, 44221 Dortmund, Germany

[§] Ioffe Institute, Russian Academy of Sciences, 194021 St. Petersburg, Russia

[∥] Collaborative Innovation Center of Extreme Optics, Shanxi University, Shanxi 030006, China

Corresponding Authors

*E-Mail: dmitri.yakovlev@tu-dortmund.de (D.R.Y.).

*E-Mail: dhfeng@phy.ecnu.edu.cn (D.F.).

ABSTRACT: Coherent spin dynamics in colloidal CdSe quantum dots (QDs) typically show two spin components with different Larmor frequencies, whose origin is an open question. We exploit the photocharging approach to identify their origin and find that surface states play a key role in the appearance of the spin signals. By controlling the photocharging with electron or hole acceptors, we show that the specific spin component can be enhanced by the choice of acceptor type. In core/shell CdSe/ZnS QDs, the spin signals are significantly weaker. Our results exclude the neutral exciton as the spin origin and suggest that both Larmor frequencies are related to the coherent spin precession of electrons in photocharged QDs. The lower frequency is due to the electron confined in the middle of the QD, and the higher frequency to the electron additionally localized in the vicinity of the surface.

**TOC GRAPHIC**:

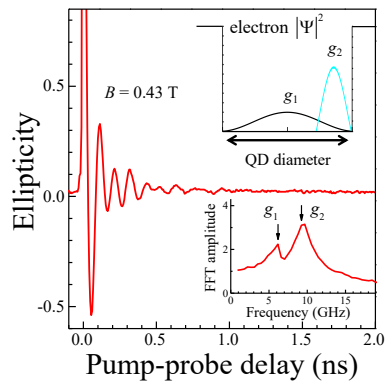

In semiconductor quantum dots (QDs) the electron wavefunction is spatially confined on the nanometer scale, which strongly modifies spin-dependent properties making most of spin relaxation mechanisms ineffective. The spins of the strongly confined electrons in QDs are considered as promising candidates for solid state quantum information processing.[1,2] Colloidal QDs have long spin coherence times[3–7] and even at room temperature the spin dephasing time in a QD ensemble is up to a few nanoseconds.[5–7] Compared with QDs grown by molecular-beam epitaxy, colloidal QDs have a much lower fabrication cost and are easier to be shape-, size- and structure-controlled. The surface of colloidal QDs being passivated with organic ligands and containing dangling bonds has a strong influence on the optical and electrical properties. Previous studies also imply that the surface affects the spin properties,[3,8,9] but these phenomena are far from being systematically investigated.

Among the colloidal nanocrystals, CdSe is the mostly investigated system in respect of photophysical processes and spin physics. The electron spin coherence in colloidal CdSe QDs has been studied by time-resolved Faraday rotation spectroscopy.[3–5,10–13] In contrast to epitaxial CdSe QDs,[14] ensembles of colloidal CdSe QDs typically show two distinct spin components with different Larmor precession frequencies and spin relaxation times. These spin signals are temperature robust and observed from liquid helium up to room temperature. The origin of the two spin components was previously assigned to the electron and exciton spins that have different $g$ factor values.[4] However, it has been demonstrated that the exciton spin relaxes fast due to strong electron−hole exchange interactions on a subpicosecond time scale at room temperature.[15,16]

In this letter, we revisit the origin of the two spin components by controlling the charge separation in CdSe QDs. We find that the charge separation selectively enhances one of the two spin components, which is dependent on the type of this separation, while enhanced charge overlap in core/shell CdSe/ZnS QDs significantly weakens both of them. These results contradict the exciton origin of one of the spin components. We suggest that the both components are related to the electrons being either confined by the QD potential or additionally localized in the surface vicinity.

A series of octadecylamine stabilized CdSe QDs in toluene (average diameter: 2.3, 2.8, 3.7, 5.6, 6.9 and 10.4 nm, estimated from the first absorption peak[17]) and CdSe/ZnS QDs (core diameter: ~4.8 nm, shell thickness: ~2.4 nm) were investigated. More information about the samples can be found in the Supporting Information. The hole acceptors 1-octanethiol (OT) and $Li[Et_3BH]$ as well as the electron acceptor 1,4-benzoquinone (BQ)[18,19] were purchased from Sigma Aldrich. OT or BQ solutions in toluene with a series of concentrations and as-grown colloidal CdSe QDs are mixed in a cuvette to realize various molar ratios of ligand to QD number, $R_{OT}$ = 70 to 70000 and $R_{BQ}$ = 100. n-type photodoping using the hole acceptor $Li[Et_3BH]$ is performed according to the method described in Ref. 19. The concentrations of CdSe QDs in each measurement are kept the same. The spin coherence dynamics are measured by two- or three-beam time-resolved ellipticity or Faraday rotation spectroscopy.[20–22] Here, circularly-polarized pump laser pulses generate carrier spin polarization in the QDs, and the subsequent spin dynamics are monitored by the change of the ellipticity or Faraday rotation of the linearly-polarized probe pulses. The time delay between the pump and probe pulses is set by a mechanical delay line. The ellipticity or Faraday rotation signals are recorded using an optical polarization bridge combined with lock-in detection. Unless otherwise stated, all measurements are performed at room temperature and in a transverse external magnetic field $B$ = 0.43 T applied in the Voigt geometry ($B$ perpendicular to the light wave vector). More experimental details can be found in the Supporting Information.

Figure 1a shows the absorption and photoluminescence (PL) spectra of as-grown CdSe QDs with a diameter of 6.9 nm. The first exciton absorption peak is at 639 nm, and the PL peak is at 653 nm. The insets of Figure 1a show the photograph of the QD solution under the illumination of ambient room light (left) and 473 nm laser (right). The corresponding ellipticity signal as a function of pump−probe delay is shown in Figure 1b with a degenerate pump−probe wavelength of 655 nm. The oscillating ellipticity signal arises from the Larmor spin precession in the external magnetic field of 0.43 T. The inset of Figure 1b is the fast Fourier transform (FFT) spectrum of the spin dynamics, which shows two Larmor precession frequencies of $v_1 = 6.21$ GHz and $v_2 = 9.53$ GHz. The two $g$ factor values of $g_1 = 1.06$ and $g_2 = 1.56$ are derived from these frequencies by the equation $g = hv_L/(\mu_B B)$, where $h$, $v_L$ and $\mu_B$ are Planck constant, Larmor precession frequency and Bohr magneton, respectively. Figure 1c shows the wavelength dependence of the FFT amplitude of spin signals in the ellipticity measurements. The Larmor precession frequency values are independent of laser wavelength near the band edge, as shown in the inset of Figure 1c. Time-resolved Faraday rotation measurements are also performed. The spin signals also contain two wavelength-independent frequencies with the same values as above. And the signal strength of both frequencies is typically weaker than that extracted from the ellipticity measurements (Figure S3).

With decreasing the QD size and increasing the optical transition energy due to stronger quantum confinement, the $g$ factor values increase, as shown in Figure 1d. Our experimental data are in good agreement with the literature data reported by Gupta et al. for CdSe colloidal QDs.[4] $g_1$ and $g_2$ are temperature independent (Figure S4), as also reported in Ref. 3. The spin dephasing time $T_2^*$ can be revealed from the width of FFT spectra. They are typically different for the two spin components being, e.g., about 380 ps for the $g_1$ component and 80−190 ps for the $g_2$ one (Figure S5). In the literature on colloidal CdSe QDs, the $g_1$ and $g_2$ components have been tentatively assigned to the electron and exciton Larmor spin precession, respectively.[4,5,10−13] These measurements have been performed on CdSe QDs with native organic ligands.

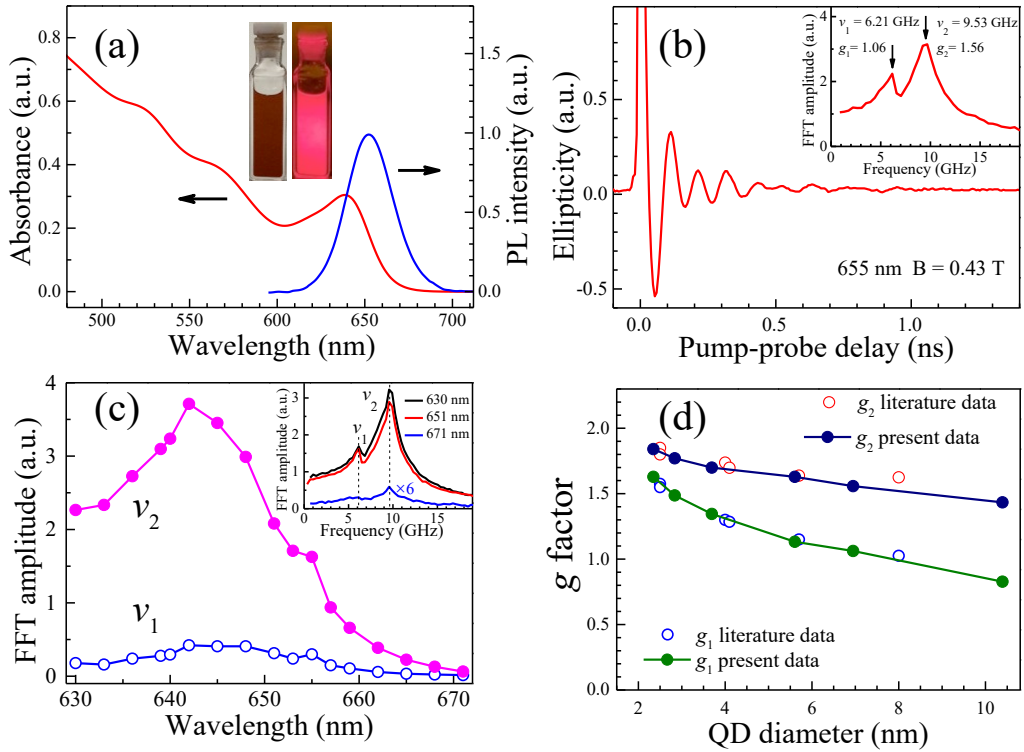

**Figure 1.** (a) Absorption and PL spectra of as-grown colloidal CdSe QDs. Insets are the QD solution under the illumination of ambient room light (left) and 473 nm laser (right). (b) Time-resolved ellipticity signal in as-grown CdSe QDs. The pump and probe wavelengths are 655 nm. (c) FFT amplitude as a function of pump−probe wavelength. The insets of panel b and c are FFT spectra. The QD diameter in panel a−c is 6.9 nm. The laser repetition rate is 30 kHz. (d) $g$ factor value as a function of QD diameter, where the literature data are taken from Ref. 4.

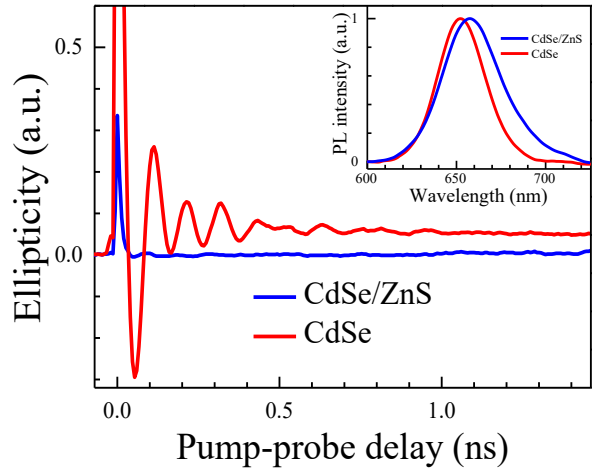

**Figure 2.** Comparison of the spin signals in as-grown CdSe QDs and core/shell CdSe/ZnS QDs. Inset: PL spectra of the two samples, both of which show a PL peak at ~655 nm. The core sizes of the as-grown CdSe QDs and core/shell CdSe/ZnS QDs are 6.9 nm and 4.8 nm, respectively. $B = 0.43$ T.

If any of the spin components originates from exciton spin precession, one would expect that the corresponding spin precession signals should readily show up in neutral QDs. Figure 2 shows a comparative measurement of as-grown bare CdSe QDs and core/shell CdSe/ZnS QDs whose PL peaks are almost at the same wavelength (see inset of Figure 2). The concentration of CdSe/ZnS QDs is adjusted and the pump−probe wavelength is set at 655 nm for the bare QDs and 650 nm for the core/shell QDs in order to have comparable laser absorption in pump−probe experiments on both samples (Figure S6). In the CdSe/ZnS QDs excitons are confined in the CdSe core and therefore isolated from the surface by the ZnS shell. The spin signal of these core/shell QDs is much weaker than that of the bare QDs (Figure 2). Furthermore, it has been demonstrated that the exciton spin in CdSe-based QDs relaxes on a subpicosecond time scale due to strong electron−hole exchange interactions at room temperature.[15,16] These relaxation times are much shorter than the spin dephasing times of 80−380 ps measured in our experiments (Figure S5). This allows us to exclude the neutral exciton as the origin of the spin signals in CdSe QDs. It consequently implies that the spin signals of both components arise from photocharged rather than neutral QDs. In the following, we manipulate the photocharging states and analyze the corresponding spin precession properties.

In the two-beam pump–probe measurements in Figure 1b with a laser repetition rate of 30 kHz, the spin amplitude of the $g_1$ component is relatively weak compared with that of the $g_2$ component. Figure 3 shows three-beam prepump–pump–probe measurement results with different prepump–pump delays implemented by changing the laser repetition rates (see the Supporting Information). One can see that the $g_1$ component can be significantly increased by introducing the third prepump pulse with linear light polarization (see the inset of Figure 3a), even when the prepump−pump delay is as large as 33.328 μs. The lifetime of the photocharging state responsible for the $g_1$ component can be evaluated by measuring the spin amplitude of the $g_1$ component as a function of prepump−pump delay $\Delta T$. The dynamics shown in Figure 3b reveal two characteristic times of 20 and 300 μs. In comparison, the spin amplitude of the $g_2$ component does not change between 1 and 30 kHz laser repetition rate (inset of Figure 3a), implying that the lifetime of the photocharging state responsible for the $g_2$ component is short compared with the laser repetition period. Note that a recent measurement of delayed exciton emission has also demonstrated a long-lived charge-separated state in CdSe QDs where the delayed emission is persistent up to a few milliseconds,[23] which might be the same state responsible for the $g_1$ component in the present work.

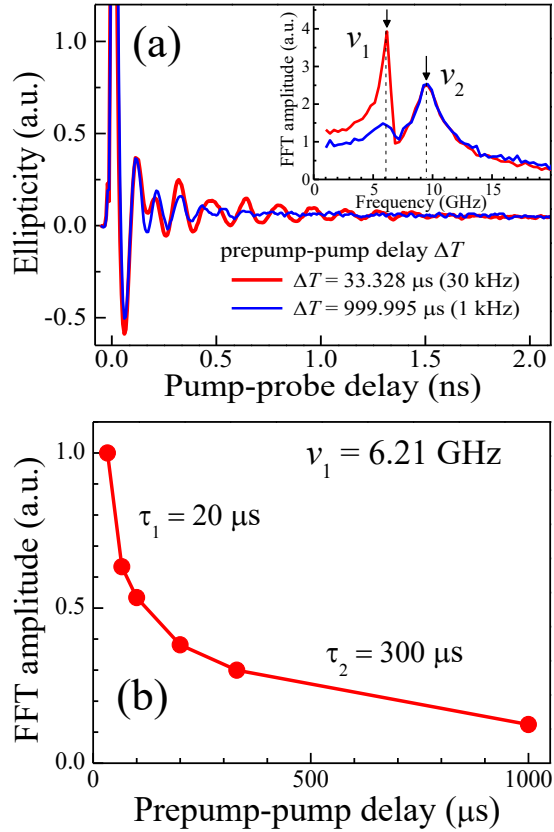

**Figure 3.** (a) Spin dynamics with different prepump−pump delays $\Delta T$ in prepump−pump−probe measurements. The pump/probe and prepump wavelengths are 655 and 455 nm, respectively. (b) FFT amplitude of the $g_1$ component as a function of prepump−pump delay. The sample is as-grown QDs with a diameter of 6.9 nm. $B = 0.43$ T.

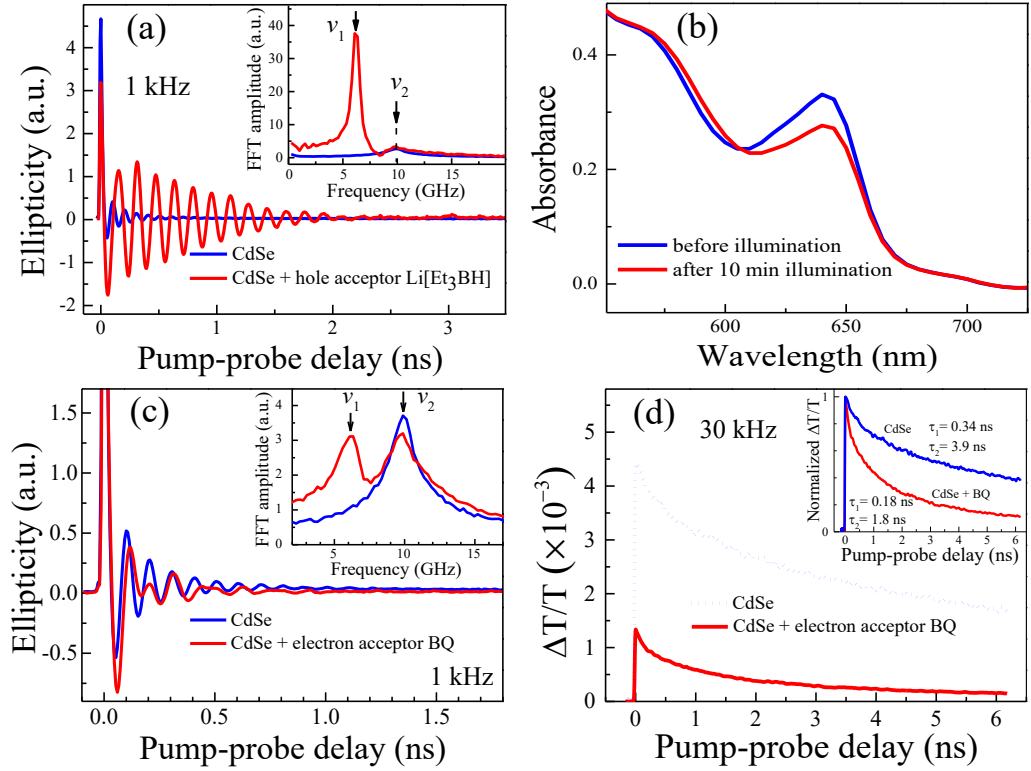

**Figure 4.** (a) Time-resolved ellipticity signals in as-grown CdSe QDs and n-type photodoped QDs using the hole acceptor $Li[Et_3BH]$. Inset shows the FFT spectra. (b) Absorption spectra of QDs with the hole acceptor $Li[Et_3BH]$ before and after light illumination. The molar ratio of $Li[Et_3BH]$ to QD in panel a and b is 120. (c) Time-resolved ellipticity signals in as-grown CdSe QDs and QDs with the electron acceptor BQ. Inset shows the FFT spectra. In panel a and c, the laser repetition rate is 1 kHz and the pump−probe wavelength is 655 nm. (d) Time-resolved differential transmission, inset presents normalized data. The pump−probe wavelength is 639 nm, set to the first exciton absorption peak. The molar ratio of BQ to QD in panel c and d is 100. The QD diameter in panel a−d is 6.9 nm.

The spin amplitude of the $g_1$ component is significantly increased by n-type photodoping in the presence of the hole acceptor Li[Et$_3$BH]. In photochemical n-doping, the photogenerated holes are captured by [Et$_3$BH]$^-$, leaving conduction band electrons in the QD core.[19] In two-beam pump−probe measurements with a laser repetition rate of 1 kHz, the $g_1$ component spin signal in the as-grown CdSe QDs is too weak to be resolvable because of the negligible pile-up effect from previous pulses. After photodoping, the spin amplitude of the $g_1$ component becomes much larger than that of the $g_2$ component as shown in Figure 4a. n-type photodoping is confirmed by the absorption bleaching at the band edge after light illumination,[19] as shown in Figure 4b.

We find that the spin amplitude of the $g_1$ component is also increased by adding the electron acceptor BQ to the QD solution and the spin amplitude of the $g_1$ component becomes comparable with that of the $g_2$ component, as shown in Figure 4c. BQ is known as electron acceptor for CdSe QDs,[18] which is confirmed by transient photobleaching measurements, as shown in Figure 4d. The photobleaching measurement is performed with a laser repetition rate of 30 kHz for a better signal-to-noise ratio. Its dynamics in the first exciton absorption peak are exclusively sensitive to the electron population in the QD.[22,24,25] In view of the nearly unchanged absorption spectrum (Figure S7a) and QD concentration, the initial electron population induced by the pump should not differ too much for the cases with and without BQ. However, the experimentally resolved initial photobleaching signal with BQ is three times weaker than that in as-grown QDs, implying that a very fast electron trapping process (on a time scale shorter than 1 ps) occurs, being beyond the experimental time resolution when BQ is added. The electron trapping may be composed of several processes. The inset of Figure 4d gives information on the two slower processes. Adding BQ, the two relaxation times become shorter and are changed from 0.34 ns and 3.9 ns to 0.18 ns and 1.8 ns, respectively. The increased relaxation rates result from the electron trapping induced by the BQ molecules. We will discuss below, that in the positively charged QDs with BQ the spin signal is formed from the electrons in the positively charged excitons ($X^+$).

The spin amplitude of the $g_2$ component is significantly increased by adding the OT hole acceptor to the QD solution. Introducing OT, the absorption spectra are almost unchanged, but the PL is quenched, as shown in Figure 5a. Figure 5b shows the Larmor spin precession in CdSe QDs with and without OT. Adding the hole acceptor ($R_{OT} = 7000$) makes the $g_2$ component three times stronger, while the $g_1$ component disappears, as shown in the inset of Figure 5b. Figure 5c shows the PL intensity and the spin amplitude as a function of the molar ratio of OT to QD. By increasing $R_{OT}$ from 0 to 70000, the PL intensity strongly decreases, while the spin amplitude of the $g_2$ component gradually increases. The PL quenching results from the hole trapping to the OT molecule at the surface.[18] The transient population of the photogenerated electrons in the QDs is nearly unaffected by the OT hole acceptor. This can be confirmed by the photobleaching dynamics in the first exciton absorption peak, as shown in Figure 5d. With and without OT, two photobleaching dynamics curves are almost the same, which is strongly different from the electron-trapping-induced photobleaching dynamics in Figure 4d.

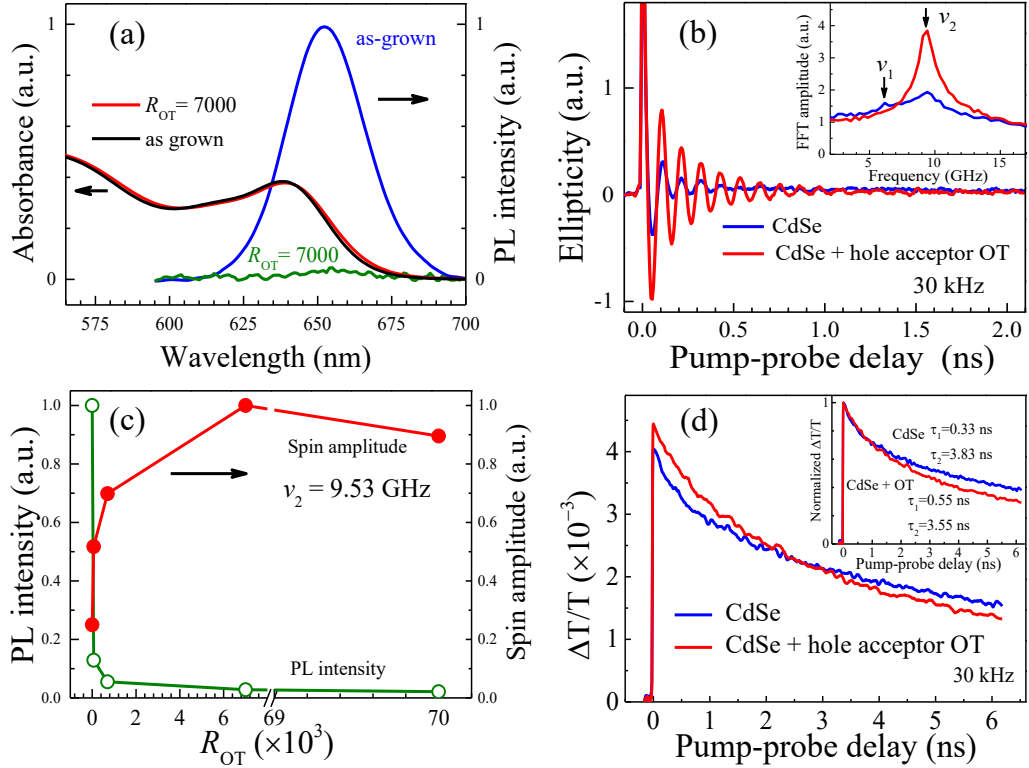

**Figure 5.** Comparison between as-grown CdSe QDs and QDs with the OT hole acceptor. The QD diameter is 6.9 nm. (a) Absorption and PL spectra. (b) Time-resolved ellipticity signals measured for the pump and probe wavelength of 655 nm. Inset: FFT spectra of these signals. (c) PL intensity and spin amplitude of the $g_2$ component as a function of $R_{OT}$. (d) Time-resolved differential transmission spectra of the two samples. The pump and probe wavelength is 639 nm, coinciding with the first exciton absorption peak. In panel a, b and d, $R_{OT}$ = 7000.

The experimental results given in Figures 4 and 5 further exclude a possible excitonic origin for any of the two spin components. Both electron and hole acceptors lead to charge separation and decrease the exciton population, while yielding a stronger spin signal of either $g_1$ or $g_2$ component. Also, if one component would be associated with a hole, the associated spin would decay fast due to the strong valence band mixing.[20] After excluding the neutral exciton and the hole as the origin of spin signals, both spin components can only originate from electron spin precession in photocharged QDs. The two distinct $g$ factor values signify that there are two types of electrons in the QDs, which differ by their wavefunction spread. As shown in Figure 1d, the variation of $g_1$ with increasing QD size tends to the value of the electron $g$ factor in bulk CdSe, $g_e = 0.68$.[26] The size dependence of the $g_1$ factor agrees well with the model calculations of the electron $g$ factor in CdSe QDs.[4,27,28] These calculations imply that the $g_1$ electron is exposed to the QD confinement potential only and, therefore, the electron wavefunction is centered in the QD.

n-type photodoping in the presence of $Li[Et_3BH]$ provides resident electrons in the QD core. Spin coherence of singly negatively charged QDs is photogenerated via the excitation of the negatively charged exciton ($X^-$) (Figure S8).[29,30] $X^-$ complex consists of two electrons with opposite spin orientations and one hole. In this case the spin is not from the optically excited complex $X^-$, but from the single, residual electron in the ground state of the electron-to-X$^-$ optical transition. Therefore, the decay of the electron spin signal is not affected by the X$^-$ recombination dynamics.

Situation is different for positively photocharged QDs with BQ. Here we also observed the increase of the spin amplitude of the $g_1$ component, but it is provided by the photogenerated electrons in the X$^+$ (Figure S9),[5] consisting of two holes and one electron. In X$^+$, the electron−hole exchange interaction characteristic for the neutral exciton is absent and does not harm the electron Larmor precession. In this case the spin signal is from the electron in X$^+$ and its decay is contributed both by the electron spin dephasing and by the X$^+$ recombination. The latter is dominated by nonradiative Auger recombination, which strongly depends on the QD size.[31,32] The FFT linewidth of the $g_1$ component in n-type photodoped QDs ($\Delta v = 0.91$ GHz) is about half of the one in positively photocharged QDs ($\Delta v = 1.95$ GHz), as shown in Figure 4a and 4c. According to $T_2^* = 1/\pi \Delta v$, the electron spin dephasing of the $g_1$ component in positively photocharged QDs is about twice faster than that in n-type photodoped QDs, implying that the nonradiative Auger process is important. In general we can conclude that $g_1$ is related to the electron in the QD center, but whether it is from negatively or positively charged QDs is dependent on the specific sample.

The spin signal of the $g_2$ component is related to the QD photocharging provided by the hole surface trapping. We conclude that from the fact that introducing hole-trapping ligands of OT increases the $g_2$ spin signal, as demonstrated in Figure 5b. The $g_2$ value is significantly larger than that of $g_1$, especially in larger QDs, which signifies that the $g_2$ electrons are subject to additional localization compared with the $g_1$ electrons. One of the possible reasons for such additional localization can be provided by the Coulomb potential of the surface-trapped holes, which would drive the electron to the surface vicinity. The photocharging state responsible for the $g_1$ spin signal is long-lived, ranging from tens microseconds in as-grown QDs (Figure 3b) to hours in photodoped QDs with $\text{Li}[\text{Et}_3\text{BH}]$ (Figure S10),[19] implying the $g_1$ electron is strongly separated from the hole. In contrast, the photocharging state responsible for the $g_2$ spin component is short-lived compared with the laser repetition periods (e.g., see the inset of Figure 3a where $g_2$ spin signals are independent of the laser repetition rate). The short lifetime of the $g_2$ electron is in accord with the supposition that the $g_2$ electron is attracted by the trapped hole.

In as-grown QDs, there are two possibilities for the excitation of the $g_2$ spin component in the two-beam pump−probe measurements. The first one is that the hole trapping rate is faster than the exciton spin relaxation rate, in which case the electron−hole exchange interaction disappears before the exciton spin relaxes completely. The second one is that the front part of the pump pulses serves as a prepump pulse resulting in a net charge separation. In comparison, the photocharging state responsible for the $g_1$ component is long-lived and the charge separated state is stored upon periodic excitation by the laser pulses (see Figure S11 and related description in the Supporting Information).

In summary, we have investigated experimentally the dynamics of the electron spin coherence in colloidal CdSe and CdSe/ZnS QDs by time-resolved ellipticity and Faraday rotation spectroscopy. These dynamics are robust even at room temperature. Two Larmor precession frequencies are found in spin signals in external magnetic fields, and the spin amplitudes depend on the type and quantity of the photocharging. We have demonstrated that both $g$ factors belong to electrons in photocharged QDs, which are either confined by the QD potential or additionally localized in the surface vicinity. Thereby, we solve the long-standing problem of the origin of the two Larmor precession frequencies in the coherent spin dynamics of colloidal CdSe QDs.

**ASSOCIATED CONTENT**

**Supporting Information**.

Sample information, details about the pump−probe experimental techniques, comparison between time-resolved ellipticity and Faraday rotation measurements, temperature dependence of spin signals, spin dephasing time as a function of QD diameter, spin excitation scheme for charged QDs, n-type photodoping with $\text{Li}[\text{Et}_3\text{BH}]$, scheme of charge excitation in prepump−pump−probe measurements as well as more supporting figures.

**AUTHOR INFORMATION**

Corresponding Authors

*E-Mail: dmitri.yakovlev@tu-dortmund.de (D.R.Y.).

*E-Mail: dhfeng@phy.ecnu.edu.cn (D.F.).

Notes

The authors declare no competing financial interest.

## ACKNOWLEDGMENTS

The authors are thankful to A. V. Rodina and Al. L. Efros for valuable discussions. This work was partially supported by the National Key Research and Development Program of China (Grant No. 2018YFA0306303), the Deutsche Forschungsgemeinschaft in the frame of the ICRC TRR 160 (Project B1), the National Natural Science Foundation of China (Grant Nos. 11374099, 11474097, 11727810 and 61720106009), the Science and Technology Commission of Shanghai Municipality (Grant Nos. 19ZR1414500 and 16520721200) and the 111 project of China (Grant No. B12024).

# Supporting Information

**1. Sample Information**

A series of octadecylamine stabilized CdSe QDs in toluene (average diameter: 2.3, 2.8, 3.7, 5.6, 6.9 and 10.4 nm) and CdSe/ZnS QDs (core diameter: ~4.8 nm, shell thickness: ~2.4 nm) were investigated. All QD samples were fabricated by Hangzhou Najing Technology Co., Ltd. The mass concentration of all obtained QD samples is 5 mg/mL. The quantum yield is >30% in CdSe QDs and >40% in CdSe/ZnS QDs. The hole acceptors 1-octanethiol (OT) and $\text{Li}[\text{Et}_3\text{BH}]$ as well as the electron acceptor 1,4-benzoquinone (BQ) were purchased from Sigma Aldrich. OT or BQ solutions in toluene with a series of concentrations and as-grown colloidal CdSe QDs are mixed in a cuvette to realize various molar ratios of ligand to QD number, where the concentrations of CdSe QDs are kept the same at 2.5 mg/mL (i.e., half of the obtained concentrations).

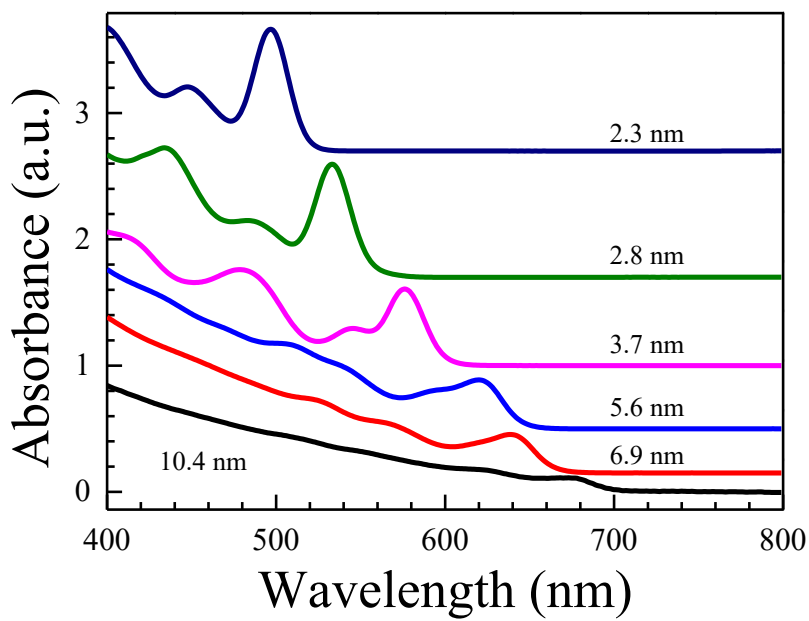

**Figure S1.** Absorption spectra of as-grown colloidal CdSe QDs with different diameters.

The QD diameters of CdSe QDs are evaluated from the first exciton absorption peak by the equation $D=(1.6122\times10^{-9})\lambda^4 - (2.6575\times10^{-6})\lambda^3 + (1.6242\times10^{-3})\lambda^2 - (0.4277)\lambda + 41.57$, where $D$ and $\lambda$ are QD diameter and wavelength of the first exciton absorption peak, respectively, both are given in nm.[1]

## 2. Pump−probe experimental techniques

Figure S2 shows the experimental configuration of two- and three-beam time-resolved Faraday rotation or ellipticity measurements. Time-resolved measurements are performed with a regenerative amplifier Yb-KGW (Ytterbium doped potassium gadolinium tungstate) laser system (PHAROS, Light Conversion Ltd.), combined with a broadband fs-OPA (optical parametric amplifier) and a narrow-band ps-OPA. In the two-beam pump−probe measurements, the pump and probe pulses are wavelength-degenerate and emitted from the ps-OPA, with their wavelength tunable around the QD bandgap. The pulse duration is ~2.5 ps and the spectral width is below 15 cm$^{-1}$. The circularly polarized pump pulses generate the spin polarization in the QD sample, while the subsequent dynamics of this spin polarization is monitored by the change of ellipticity or Faraday rotation of the linearly polarized probe pulses. In time-resolved ellipticity measurements, the linearly polarized probe light will become partly elliptically polarized because of the absorption difference of left and right circularly polarized light in the spin-polarized system. Similarly, in time-resolved Faraday rotation measurements, the plane of polarization of the probe light will be rotated due to the refraction index difference of the two circularly polarized components. In the three-beam prepump−pump−probe measurements, the prepump pulse is linearly polarized with the duration of 200 fs, taken from the fs-OPA. The prepump pulses are used to induce photocharging in the QDs. The repetition rate of both OPAs is 30 kHz, but it can be divided by integers via a pulse picker inside the amplifier without changing the pulse energy.

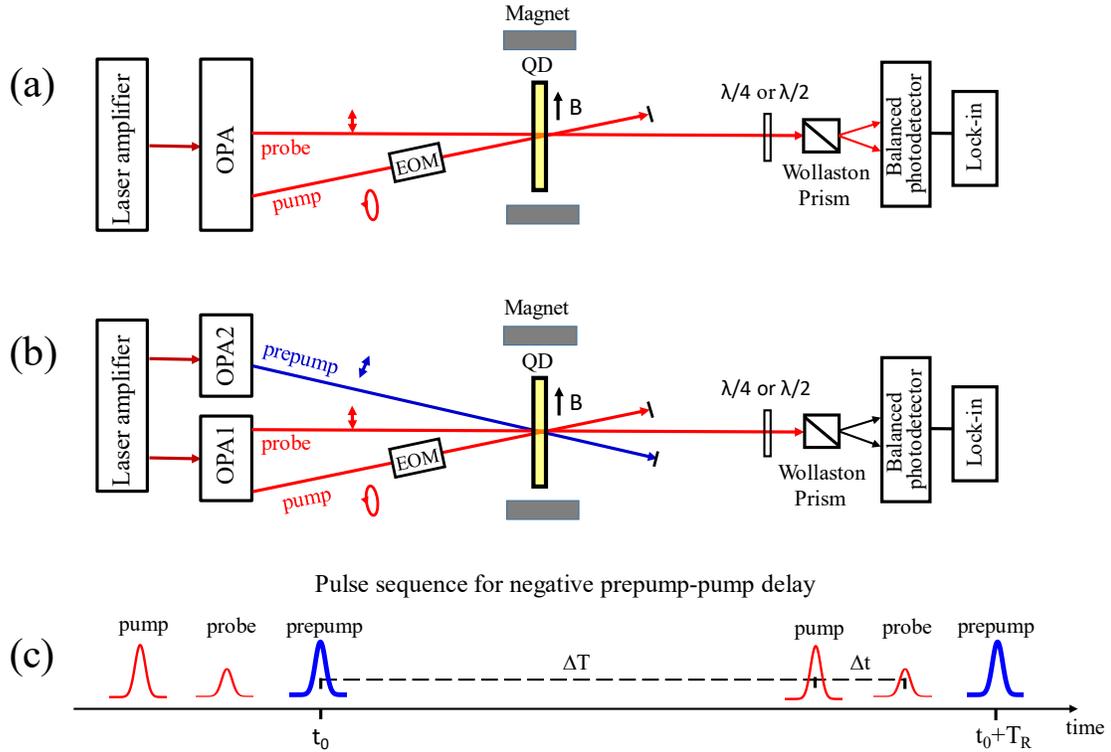

**Figure S2.** Experimental configuration. (a) Two-beam time-resolved Faraday rotation or ellipticity measurements. (b) Three-beam time-resolved Faraday rotation or ellipticity measurements. (c) Pulse sequence for negative prepump–pump delay. For the 30 kHz laser repetition rate, $\Delta T = 33.328$ μs is equivalent to $\Delta T = -5$ ns. $T_R$ is the laser repetition period.

In the prepump–pump–probe measurements, a long prepump–pump delay time of several tens μs to ms is obtained with a pump–probe–prepump pulse sequence, i.e., a negative prepump–pump delay as shown in Figure S2c. For 30 kHz laser repetition rate, the prepump–pump delay of $\Delta T = 33.328$ μs is equivalent to $\Delta T = -5$ ns. For 1 kHz laser repetition rate, the same negative delay is corresponding to the prepump−pump delay of $\Delta T = 999.995$ μs.

In the time-resolved differential transmission measurements, both the pump and probe pulses are linearly polarized, and their wavelengths are set at the first exciton absorption peak. The pump and probe beams are intensity-modulated at different frequencies via the inner and outer circle of slots of an optical chopper, respectively, and the signal is detected at their sum frequency.

**3. Wavelength dependence of spin signals in time-resolved ellipticity and Faraday rotation measurements**

Figure S3 shows the wavelength dependence of the spin amplitudes extracted from the ellipticity and the Faraday rotation measurements. Both Larmor precession frequency values are independent of laser wavelength near the band edge, as shown in the inset of Figure S3. The pump/probe wavelength only changes the relative amplitude of spin signals. Similar phenomena were also found in the previous literature.[2] The reason making the *g* values independent of the laser wavelength is beyond our present research. In our measurements, the ellipticity signal has typically larger amplitude than the Faraday rotation signal. In contrast to the ellipticity signals, the Faraday rotation signals show a dispersive line shape near 650 nm with changing phase by π.

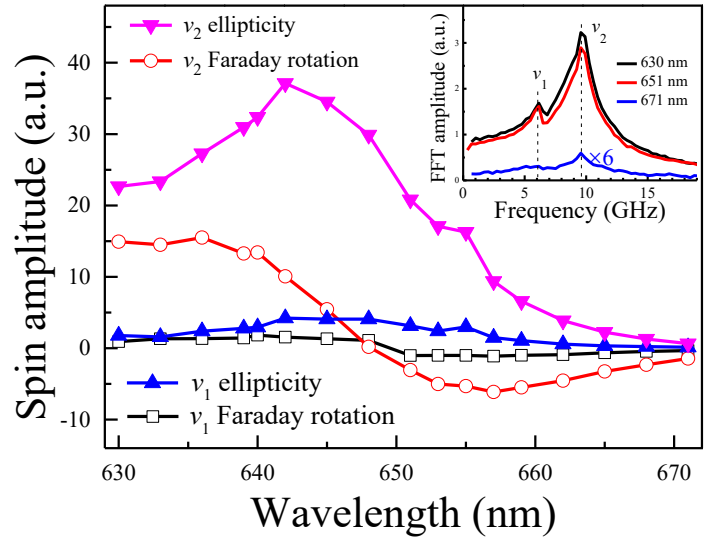

**Figure S3.** Wavelength dependence of spin amplitudes in time-resolved ellipticity and Faraday rotation measurements for as-grown CdSe QDs with a diameter of 6.9 nm. Inset: FFT spectra of these signals. $v_1$ = 6.21 GHz and $v_2$ = 9.53 GHz. Note, that the negative sign of the spin amplitude means that the spin signal phase is changed by $\pi$.

**4. Temperature dependence of spin coherence dynamics**

Figure S4 shows time-resolved Faraday rotation signals and their FFT spectra in CdSe QDs at different temperatures. The two *g* factor values remain unchanged in the temperature range from 3 K to 300 K. The relative intensity of the two spin components varies for different temperatures. Discussion of this variation is beyond the scope of the present paper.

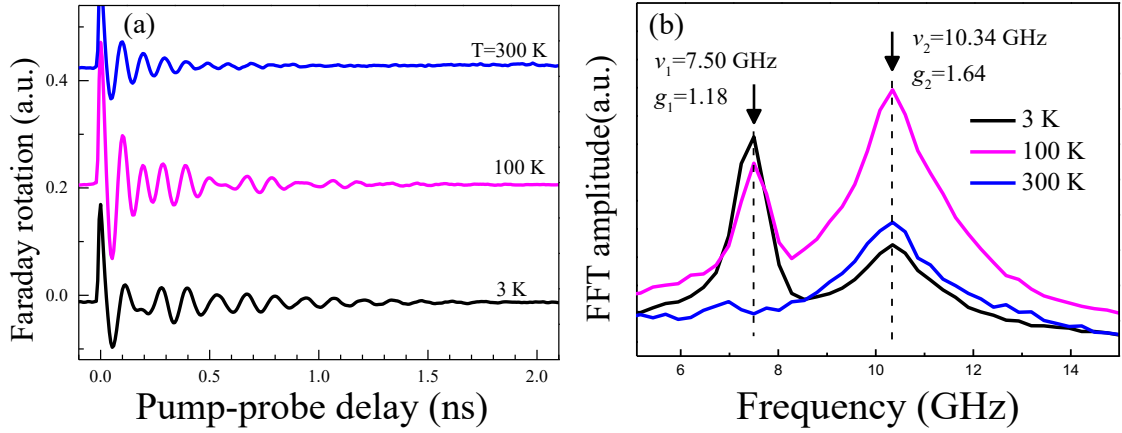

**Figure S4.** (a) Faraday rotation signals for as-grown CdSe QD film as a function of pump−probe delay at different temperatures. (b) FFT spectra of panel a. The QD size is ~5.9 nm. $B = 0.5$ T.

## 5. Spin dephasing time as a function of QD diameter

The spin dephasing times $T_2^*$ extracted from the width of FFT spectra are different for the two spin components. Figure S5 shows the QD size dependence of $T_2^*$ extracted from the full width at half maximum of the FFT spectra ($\Delta v$) for the $g_1$ and $g_2$ components, using the equation $T_2^*=1/\pi \Delta v$. The spin dephasing time of the $g_1$ component ($T_{2,g_1}^*$) is longer than that of the $g_2$ component ($T_{2,g_2}^*$). Decreasing the QD diameter from 6.9 nm to 2.3 nm, $T_{2,g_2}^*$ increases from 80 ps to 190 ps, while $T_{2,g_1}^*$ remains almost constant at about 380 ps. Various factors may affect the spin dephasing time, such as inhomogeneous broadening, electron−nuclear hyperfine interaction, quantum confinement, surface effects, and carrier lifetime. The reason for the two different dephasing times is complicated and goes beyond the scope of this paper. Note that the *g* factor value and spin dephasing time for the dot diameters of 2.3 and 2.8 nm are extracted from the time-resolved ellipticity measurements of CdSe QDs with OT molecules,[3] as the spin signals in as-grown QDs are too weak to give useful information.

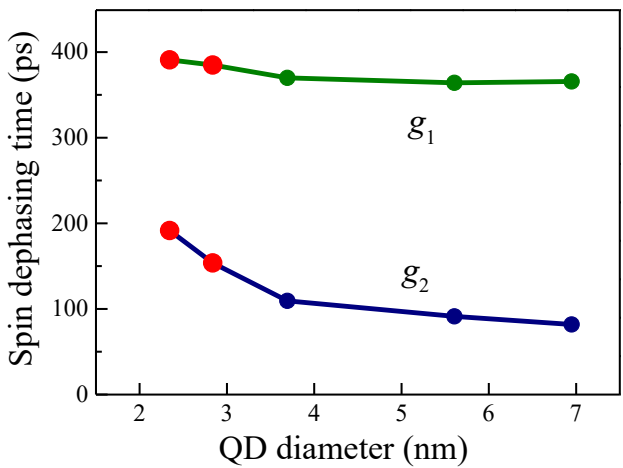

**Figure S5.** Spin dephasing time $T_2^*$ as a function of CdSe QD diameter. $T = 300$ K, $B = 0.43$ T. The red point data are obtained from CdSe QDs with OT molecules.[3]

## 6. Absorption and PL spectra in bare CdSe and core/shell CdSe/ZnS QDs

Figure S6 shows absorption and PL spectra of bare CdSe and core/shell CdSe/ZnS QDs. The diameter of the bare CdSe QDs is 6.9 nm. The core diameter, shell thickness and total diameter of the CdSe/ZnS QDs are 4.8 nm, 2.4 nm and 9.6 nm, respectively. The first exciton absorption peak in the bare CdSe QDs is at 639 nm, and the PL peak is at 653 nm. The first exciton absorption peak in the core/shell CdSe/ZnS QDs is at ~645 nm, and the PL peak is at 658 nm. Despite the type-I band alignment in CdSe/ZnS QDs, adding a ZnS shell will lead to a red shift of the absorption and PL spectra due to the penetration of the electron wavefunction into the ZnS shell.[4] There are two choices for the sample selection: (a) The same core size for bare and core/shell structures, but that requires different laser conditions (the first absorption peak of 4.8 nm bare CdSe QDs is ~605 nm, while it is ~645 nm in CdSe/ZnS QDs). In our OPA system, 640 nm is from the signal photons, while 605 nm from the second harmonic of idle photons. Wavelength change thus increases the experiment complexity and result uncertainty. (b) Different core size but with close absorption and PL wavelengths, is preferred in our measurements. Although the core size is different, it is not matter because 4.8 and 6.9 nm CdSe QDs have similar spin phenomenon. Although the obtained bare CdSe and core/shell CdSe/ZnS QDs have the same mass concentration, the molar concentration of these two samples are different due to the mass difference between the bare core and core/shell QDs. The concentration of the CdSe/ZnS QDs is adjusted, and the pump−probe wavelength is set at 655 nm for the bare dots and 650 nm for the core/shell dots in order to have comparable laser absorption in the pump−probe experiments.

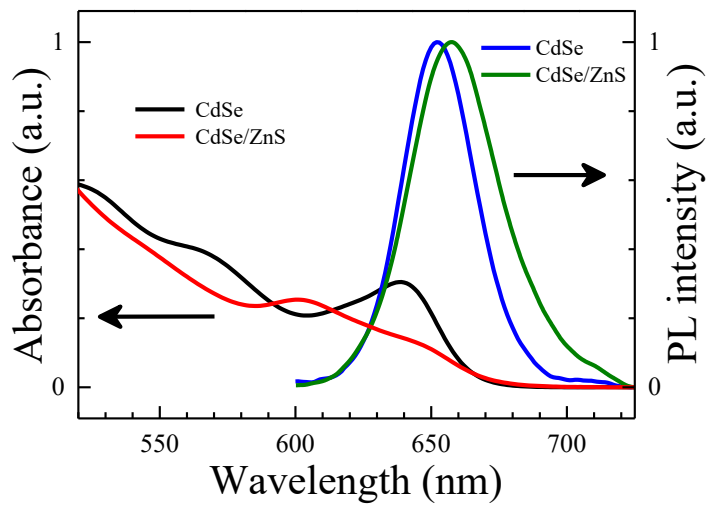

**Figure S6.** Absorption and PL spectra of bare CdSe and core/shell CdSe/ZnS QDs. The core size of the as-grown CdSe QDs is 6.9 nm. The core diameter, shell thickness and total diameter of the CdSe/ZnS QDs are 4.8 nm, 2.4 nm and 9.6 nm, respectively.

## 7. Steady state absorption and PL spectra for CdSe QDs with and without electron acceptor BQ

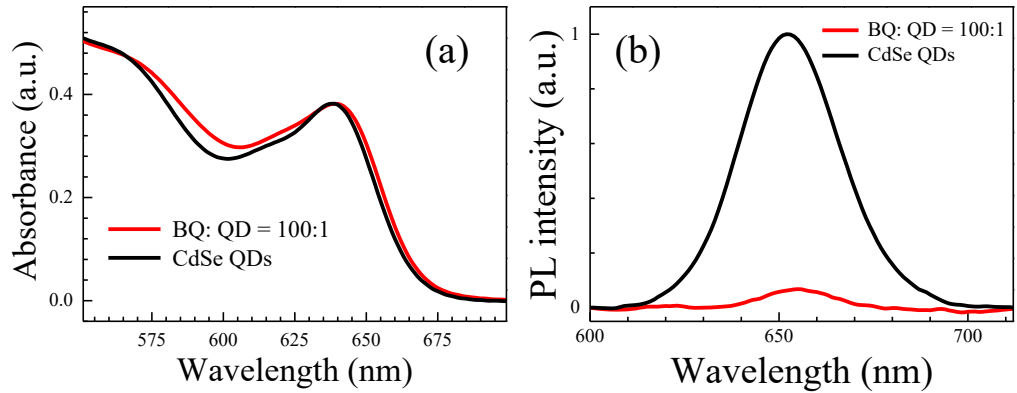

**Figure S7.** (a) Band edge absorption spectra are almost the same for CdSe QDs with and without the electron acceptor BQ. (b) Adding the electron acceptor BQ quenches the QD PL intensity. The diameter of the CdSe QDs is 6.9 nm.

## 8. Spin excitation scheme for negatively charged QDs

Figure S8 shows the scheme of spin excitation in negatively charged QDs. The spin signal results from polarization-selective electron-to-negative trion excitation. The negative trion singlet ground state consists of two electrons with opposite spin orientations and a single hole with spin ±3/2, with the total spin defined by the hole spin. Optical selection rules only allow +1/2 (-1/2) electrons to be excited into the +3/2 (-3/2) trion states by $\sigma^+(\sigma^-)$ circularly polarized laser pulses, leaving a net part of spin-down (spin-up) polarized electrons in the ground states.

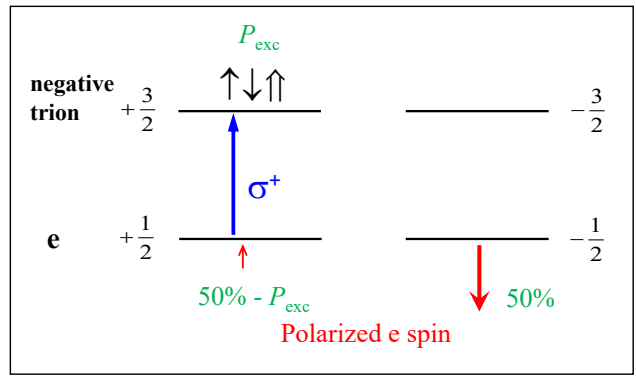

**Figure S8.** Scheme of spin excitation in negatively charged QDs

## 9. Spin excitation scheme for positively charged QDs

Figure S9 shows the scheme of spin excitation in positive charged QDs. The spin signal comes from polarization-selective hole-to-positive trion excitation. The positive trion singlet ground state consists of two holes with opposite spin orientations and a single electron with spin $\pm 1/2$, where the total spin is defined by the electron spin. According to the optical selection rules, $\sigma^+(\sigma^-)$ circularly polarized pulses can only excite -3/2 (+3/2) holes to -1/2 (+1/2) trion states and generate spin-down (spin-up) polarized electrons in the positive trion states. Note that there are no electron−hole exchange interactions in the ground singlet states of both positive and negative trions. For both negatively and positively charged QDs, any hole spin polarization involved in the excitation will be not resolved in our measurements, as the hole spin decays very fast at room temperature.

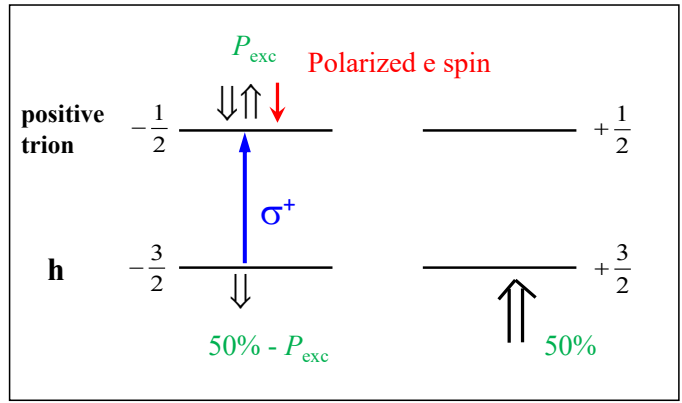

**Figure S9.** Scheme of spin excitation in positively charged QDs

## 10. n-type photodoping with Li[Et$_3$BH][5]

Before the light illumination for photodoping the spin signal is very weak (the lowest black curve in Figure S10a) and it belongs to the $g_2$ component as revealed from the FFT spectrum shown in Figure S10b. After switching on the photodoping light for 10 min, the spin signal increases remarkably. The enhanced spin signal belongs to the $g_1$ component. The charge separation state that corresponds to the $g_1$ component is long-lived. As shown in Figure S10a, after 10 min illumination and switching off the photodoping light, the spin signal of the $g_1$ component remains strong. This demonstrates that the charging state responsible for the $g_1$ component lives more than 60 min.

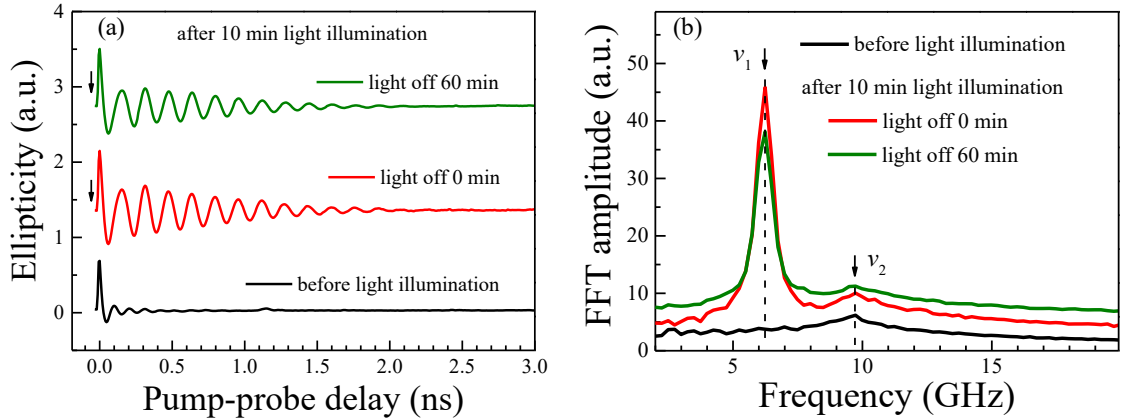

**Figure S10.** (a) Time-resolved ellipticity signals for different illumination conditions of the photodoping light. We use a femtosecond laser for photodoping with a wavelength of 655 nm, the laser spot size of 10 mm and the power of 10 mW. The black arrows on the left side denote the start point for counting the photodoping-light off time. (b) FFT spectra of the signals in panel a. The molar ratio of $Li[Et_3BH]$ to QD is 120. The QD diameter is 6.9 nm, $B$ = 0.43 T. The plots are offset in order to better distinguish the two different frequencies.

## 11. Scheme of charge excitation in prepump−pump−probe measurements

Figure S11 shows the scheme of charge excitation related to $v_1$ component in as-grown QDs in prepump−pump−probe measurements. The linearly polarized prepump pulse generates the electron−hole pairs in the QD. After some time, the QD becomes negatively charged due to the hole trapping at the QD surface. Once the QD becomes negatively charged, a circularly polarized pump pulse will polarize the electron spin (see Figure S8). In the measurements, the prepump could be even applied after the pump and probe pulses if the charging lifetime is very long. In this case, the former prepump pulses are served for generating charged states. Even without the prepump, the former pump pulses can also generate charged states whose quantity depends on the laser fluence, wavelength and repetition rates.

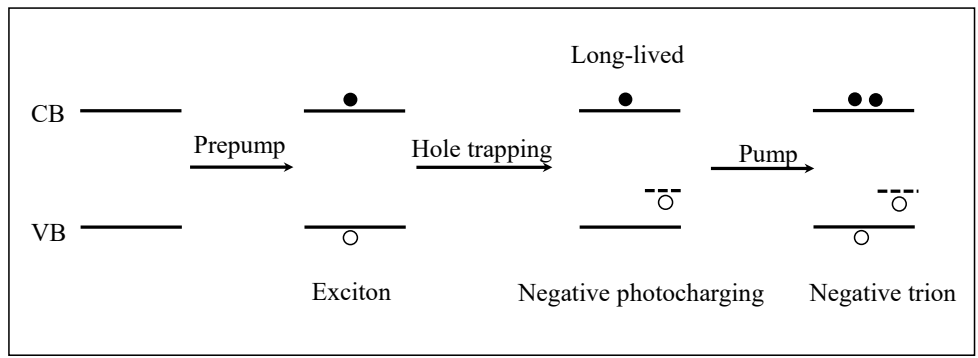

**Figure S11.** Scheme of charge excitation in prepump–pump–probe measurements.